\documentclass[draft]{agujournal2018}
\usepackage{apacite}
\usepackage{url} 
\usepackage{lineno}
\usepackage[symbol]{footmisc}  
\renewcommand{\thefootnote}{\fnsymbol{footnote}}

\journalname{Geophysical Research Letters}

\begin{document}


\title{Morphology and dynamics of Venus's middle clouds with Akatsuki/IR1\footnote[1]{Original here: https://agupubs.onlinelibrary.wiley.com/doi/full/10.1029/2018GL081670}.}

%
%

\authors{J. Peralta\affil{1},
         N. Iwagami\affil{2},
	       A. S\'anchez-Lavega\affil{3},
		     Y. J. Lee \affil{4},
		     R. Hueso \affil{3},
		     M. Narita \affil{4},
		     T. Imamura \affil{4},
	       P. Miles \affil{5},
		     A. Wesley \affil{6},
	       E. Kardasis\affil{7} and
		     S. Takagi\affil{8}
				 }

\affiliation{1}{Institute of Space and Astronautical Science (ISAS), Japan Aerospace Exploration Agency (JAXA), 3-1-1, Yoshinodai, Chuo-ku, Sagamihara, Kanagawa, 252-5210, Japan}
\affiliation{2}{Tokyo 156-0044, Japan}
\affiliation{3}{Escuela de Ingenier\'ia de Bilbao, University of the Basque Country (UPV/EHU), Bilbao, Spain}
\affiliation{4}{Graduate School of Frontier Sciences, The University of Tokyo, Japan}
\affiliation{5}{Rubyvale QLD, Australia}
\affiliation{6}{Astronomical Society of Australia, Australia}
\affiliation{7}{Hellenic Amateur Astronomy Association, Greece}
\affiliation{8}{Research and Information Center, Tokai University, Japan}

\correspondingauthor{Javier Peralta}{javier.peralta@jaxa.jp}

\begin{keypoints}
\item First extensive study (more than a year) of the middle clouds of Venus at low latitudes combining Akatsuki and ground-based observations.
\item Cloud morphologies observed at high spatial resolution and with high contrasts suggest important differences between middle and upper clouds.
\item Middle cloud winds peak at the equator and have long-term variations when compared with results from previous missions.
\end{keypoints}

%

\begin{abstract}
The Venusian atmosphere is covered by clouds with super-rotating winds whose accelerating mechanism is still not well understood. The fastest winds, occurring at the cloud tops ($\sim$70 km height), have been studied for decades thanks to their visual contrast in dayside ultraviolet images. The middle clouds ($\sim$50-55 km) can be observed at near-infrared wavelengths (800-950 nm), although with very low contrast. Here we present the first extensive analysis of their morphology and motions at lower latitudes along 2016 with 900-nm images from the IR1 camera onboard Akatsuki. The middle clouds exhibit hemispherical asymmetries every 4--5 days, sharp discontinuities in elongated ''hook-like'' stripes, and large contrasts (3-21\%) probably associated with large changes in the optical thickness. Zonal winds obtained with IR1 images and with ground-based observations reveal mean zonal winds peaking at the equator, while their combination with Venus Express unveils long-term variations of 20 m s$^{-1}$ along 10 years.
\end{abstract}

%
%

\section{Introduction}\label{sec:intro}
Venus's clouds are mainly composed of H$_2$SO$_4$ and aerosols of three main sizes or modes (1, 2 and 3) \citep{Esposito1983,McGouldrick2012}. The clouds form an extended vertical layer ($\sim$50--70 km) that permanently covers the planet and is divided into three decks --upper, middle and lower clouds-- shrouded by haze \citep{Knollenberg1980clouds}. At the upper and middle clouds of Venus, the zonal superrotation exhibits the largest speeds and vertical shear \citep{Sanchez-Lavega2017} and is also where most of the solar energy is deposited \citep{Titov2013}. The upper clouds' top at $\sim$70 km \citep{Ignatiev2009} can be observed with ultraviolet (UV) and violet wavelengths ($\sim$360--480 nm), with their morphology and dynamics been extensively studied for decades \citep{Belton1976a,Rossow1980,Rossow1990,Titov2012,Khatuntsev2013,Limaye2018,Horinouchi2018} thanks to the strong contrasts caused by an unknown absorber \citep{Lee2015Icar,Perez-Hoyos2018}.\\

\null
Since photons with longer wavelengths can penetrate the Venusian clouds deeper before being reflected \citep{Sanchez-Lavega2008,Takagi2011}, the middle-lower clouds of Venus can be observed on the dayside at 570--680 and 900--1000 nm, although with weaker contrast \citep{Belton1991,Hueso2015}. The altitude of these contrasts is not well constrained and different estimates have been obtained with radiative transfer calculations (51--55 km from \citealt{Iwagami2018} based on \citealt{Takagi2011}; 58--68 km from \citealt{Khatuntsev2017}) or comparing cloud-tracked speeds with vertical profiles of the wind from entry probes (55 km from \citealt{Belton1991}; 62 km from \citealt{Peralta2007b}; 50--57 km from \citealt{Khatuntsev2017}). First studies of these clouds came from polarized images at 935 nm during the Pioneer Venus mission \citep{Limaye1984}, although it was not until the analysis of Galileo in 1990 that their different morphology and slower wind speeds became evident through 986-nm images \citep{Belton1991,Peralta2007b}. Wind measurements during the Venus Express (VEx) mission and the MESSENGER's flyby confirmed these results, and also showed that middle clouds' winds can be rather variable \citep{Sanchez-Lavega2008,Hueso2012,Hueso2015,Peralta2017GRL,Khatuntsev2017}. More recently, the middle clouds have been observed by amateur astronomers \citep{Mousis2014,Sanchez-Lavega2016,Kardasis2017} and with higher detail with JAXA's Akatsuki mission \citep{Iwagami2018,Limaye2018}.\\

\null
Here we report on the first year (2016) of Akatsuki observations of the middle clouds of Venus on its dayside with the 900-nm images obtained with the 1-$\mathrm{\mu m}$ camera (h.a. IR1), complemented with images at similar wavelengths obtained with small telescopes from Greece and Australia in coordination with the Akatsuki mission (also ``Venus Climate Orbiter'' or VCO). A description of the images used in this work and the techniques to process them and measure wind speeds are included in section \ref{sec:methods}. The new cloud morphologies in images from IR1 and their contrasts are discussed in sections \ref{sec:clouds} and \ref{sec:albedo}. Finally, the wind results and their long-term behaviour are discussed in section \ref{sec:winds}.

\section{Methods}\label{sec:methods}
We inspected the full set of 984 dayside images from IR1 (calibration version "v20170601''), covering from 2015 December 7 to 2016 December 9, when both IR1 and 2-$\mathrm{\mu m}$ (IR2) cameras ceased their observations \citep{Satoh2017,Iwagami2018}. The IR1 dataset\footnote{Available at: \url{http://darts.isas.jaxa.jp/planet/project/akatsuki/}} was complemented by ground-based observations with small telescopes using high-resolution amateur techniques \citep{Mousis2014,Sanchez-Lavega2016}, particularly useful when the observation phase angle from Akatsuki was large. Nineteen images were acquired from Rubyvale Observatory (Australia) using a 508-mm F4 Newtonian telescope, a CMOS camera and a 1.0--1.1 $\mathrm{\mu m}$ filter. These images\footnote{Available at: \url{http://www.acquerra.com.au/astro/gallery/venus/index.live}} covered 13 days from 2016 October 6 to November 9, when Venus had a solar elongation of 35$^{\circ}$, apparent angular diameter of about 13'', and $\sim$80\% of illuminated fraction. Venus was also observed from Dimitra Observatory in Glyfada-Athens (Greece) during 2016 December 15 and 2017 January 2, 3 and 13. Ten images\footnote{Available at: \url{http://kardasis.weebly.com/eastelong-2016.html}} were taken using a 14-inch Celestron telescope and Hutech 884--900 nm bandpass filter, with Venus presenting solar elongation of $\sim$80$^{\circ}$, $\sim$21'' diameter and $\sim$56\% of illuminated disk.\\

\null
Due to the low contrast ($\sim$4\%) of Venus within 900--1000 nm (h.a. \textit{near-infrared} or \textit{NIR}) \citep{Belton1991}, IR1 images were selected attending to the spatial resolution, signal-to-noise ratio and phase angle, prioritizing the presence of cloud features easy to track. Uncertainties in the navigation of IR1 images was corrected using an automatic ellipse fitting \citep{Ogohara2017}, while for the navigation and adjustment of size/position/orientation of the grid of ground-based images we used WinJupos \citep{Hahn2012} or, alternatively NASA's SPICE kernels \citep{Acton1996} and the interactive software created by \citet{Peralta2018ApJS}.

\subsection{Wind measurements}\label{ssec:winds}
The images selected to infer winds (83 images, $\sim$8\% of the dataset) were processed as it follows:
\begin{enumerate}
	\item \textit{Image processing}: we improved the visualization with a Minnaert photometric correction (see subsection \ref{ssec:photometry}) followed by an unsharp-mask to enhance subtle features and brightness/contrast adjustment.
	\item \textit{Geometrical projections}: pairs of IR1 images were projected onto equirectangular (cylindrical) geometry, using an angular resolution consistent with the original images. Since Akatsuki's equatorial orbit favors observing middle-to-lower latitudes, only pericenter images had enough spatial resolution to infer high-latitude winds from azimuthal equidistant (polar) projections. The angular resolution for polar projections was that of original images at $\sim$70$^{\circ}$ latitude. Regarding ground-based images, cylindrical projections with angular resolution ranging 5$^{\circ}$--8$^{\circ}$ per pixel were used (depending of Venus's disk size and assuming a seeing of $\sim$0.6''). The spatial resolution of the projections varies from 16--54 km per pixel (IR1) to $\sim$530--850 km (ground-based).
	\item \textit{Wind measurements}: winds were measured using cloud tracking in pairs of images. A \textit{semi-automatic} method \citep{Peralta2018ApJS} was employed, using \textit{phase-correlation} for the template matching between images and the final result being accepted/rejected by a human operator. When the \textit{phase-correlation} was not effective, the classical visual method undertaken by an operator was used \citep{Peralta2007b,Sanchez-Lavega2008}.
\end{enumerate}

\subsection{Photometric correction}\label{ssec:photometry}
We used 644 IR1 images solar-target-observer phase angle or simply \textit{solar phase angle} ranging 2$^{\circ}$--154$^{\circ}$, 65\% of the full dataset) to evaluate the best photometric correction for dayside images of Venus at 900 nm. Instead of the \textit{radiance factor} \citep{Lee2017}, we used the \textit{observed radiance} (mW m$^{-2}$sr$^{-1}\mathrm{\mu m}^{-1}$) as $I_{obs}(\alpha,\mu,\mu_{0})\propto D(\alpha,\mu,\mu_{0})$, where $D$ is a disk function describing the photometric angle dependence, $\alpha$ is the solar phase angle (degrees), $\mu=cos(e)$ and $\mu_{0}=cos(i)$, with $e$ being the emergence angle and $i$ the incidence angle. To evaluate $D$, we compared \textit{Minnaert} and \textit{Lambert-Lommel-Seeliger} laws as \citet{Lee2017}, in this case with $i<$84$^{\circ}$ and $e<$84$^{\circ}$. The \textit{Minnaert} law ($D_{Mi}=\mu_{0}^{k_{Mi}(\alpha)}\mu^{k_{Mi}(\alpha)-1}$) was found to perform better, with the coefficient $k_{Mi}(\alpha)$ fitting the polynomial function:
\begin{equation}
	k_{Mi}(\alpha)=0.932486+0.00673108\cdot\alpha-4.75720\cdot 10^{-5}\cdot\alpha^{2}-1.11378\cdot10^{-8}\cdot\alpha^{3}
\label{eq:Minnaert}
\end{equation}

\section{Cloud morphologies at 900 nm}\label{sec:clouds}
\begin{figure*}[h!]
\centerline{\includegraphics[width=140mm]{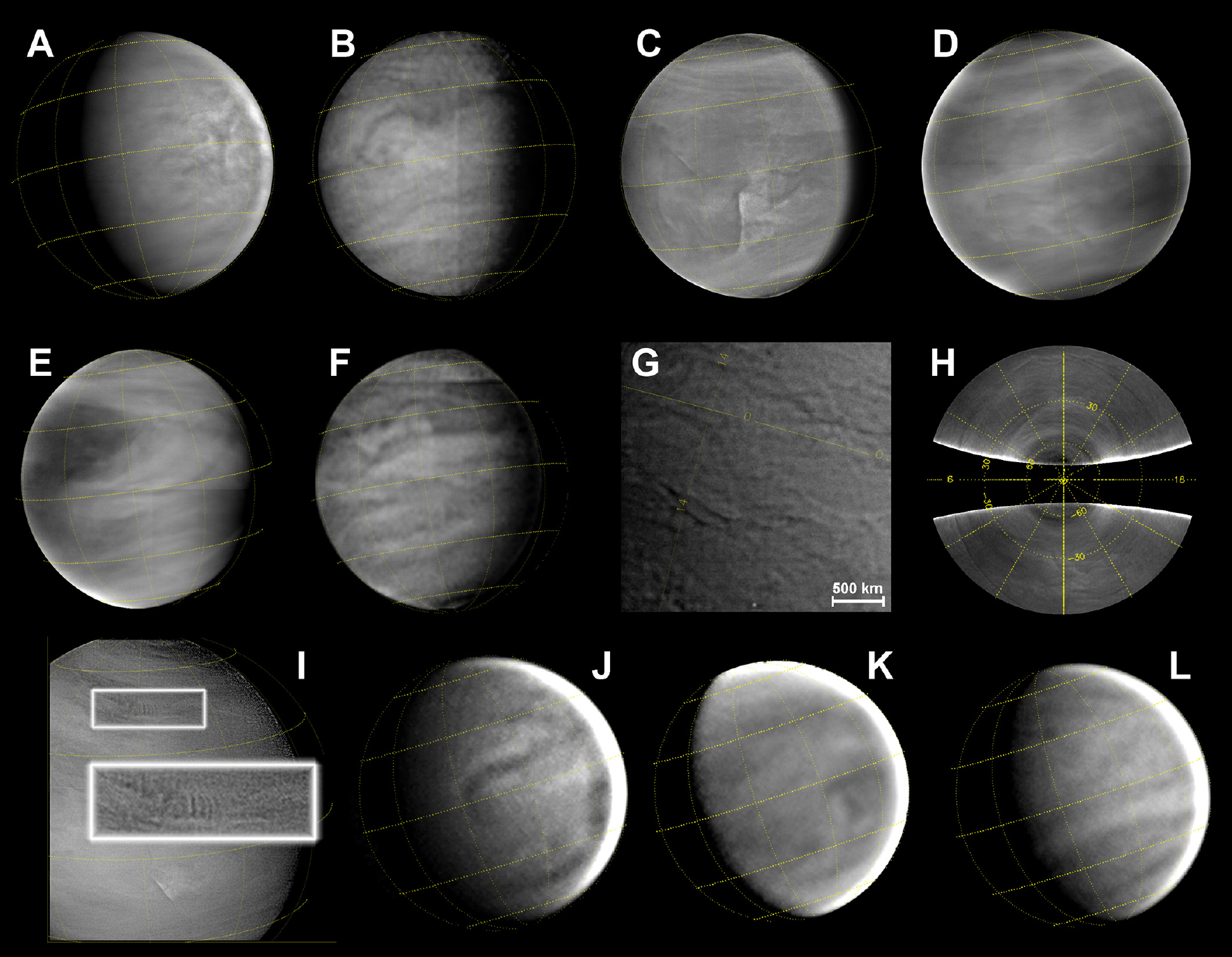}}
\caption{\textbf{The variable morphology of the middle clouds.} Examples of cloud patterns on the albedo at 900 nm from Akatsuki/IR1 (A--I) and ground-based images (J--L) are shown. Grids with latitudes 0$^{\circ}$, 30$^{\circ}$ and 60$^{\circ}$ are over-plotted with yellow contours. Image H exhibits polar projections of the north (above) and south (below) hemispheres of D. Image I shows a wave packet on the northern hemisphere and a zoom of it. Images A,C,D,E and I have a spatial resolution of $\sim$18 km per pixel (Akatsuki located at $\sim$90,000 km from Venus), B and F have a resolution of $\sim$71 km per pixel ($\sim$350,000 km), and G is a close-up image (18,000 km) showing fine scales at a spatial resolution of 2.5 km per pixel. The ground-based images (J--L) contain details down to a scale of $\sim$500 km. The images were taken at dates: (A) 2015-12-07T08:32:01, (B) 2016-05-02T15:02:07, (C) 2016-04-25T17:02:07, (D) 2016-05-06T16:02:09, (E) 2016-05-17T18:02:07, (F) 2016-05-03T15:02:07, (G) 2016-09-16T01:27:08, (H) 2016-05-06T16:02:09, (I) 2016-11-20T15:05:52, (J) 2016-10-05T07:25:39, (K) 2016-10-10T07:34:58, and (L) 2016-10-17T07:44:46. All images were processed as explained in subsection \ref{ssec:winds}.}
\label{figure:IR1_clouds}
\end{figure*}

Figure \ref{figure:IR1_clouds} exhibits examples of the NIR albedo of Venus's clouds during 2016, showing unseen morphologies with noticeable changes along time. When observed in UV, low and middle latitudes seem dominated by a dark equatorial band with mottled and patchy appearance \citep{Belton1976a,Rossow1980,Titov2012,Sanchez-Lavega2016}. NIR images also display a slightly darker band though normally invaded by bright clouds, which sometimes have swirl-shape and mottled aspect (Figs.~\ref{figure:IR1_clouds}A--C and G) similar to those seen during MESSENGER's flyby \citep{Peralta2017GRL} and VEx mission \citep{Markiewicz2007a}, and suggestive of convection. Other times, the previous turbulent regime seems to evolve to a laminar one, with clouds becoming homogeneously bright and/or featureless, conforming multiple stripes with quasi-zonal orientation (Fig.~\ref{figure:IR1_clouds}D--F). When this apparent laminar regime seems dominant, symmetry is observed between the north and south polar regions (Fig.~\ref{figure:IR1_clouds}H). Clouds' bands have lengths of 2,000-4,000 km and they are tilted relative to the parallels except for those at $\sim$11$^{\circ}$S during October 17 with width of 700 km and length $>$11,000 km (Fig.~\ref{figure:IR1_clouds}L). From April to May, the northern hemisphere up to 45$^{\circ}$N became periodically darkened (radiance decrease of $\sim$5\%) every 4--5-days (Figs.~\ref{figure:IR1_clouds}E--F). Such strong hemispherical asymmetries have never been reported on the UV albedo, and their cause is yet to be determined although inhomogeneities in the distribution of an absorber at NIR wavelengths \citep{Titov2012} cannot be ruled out. The sharp albedo changes displayed in Fig.~\ref{figure:IR1_clouds}C have never been observed before Akatsuki \citep{Limaye2017}, and they were recurrent during the first half of 2016 and absent on UV images of the upper clouds. They involve a 1\%--4\% decrease in the photometrically-corrected radiance and, since they moved with zonal speeds $\sim$10--20 m s$^{-1}$ faster than other cloud features, we interpret them as atmospheric waves.\\

\null
Other cloud features are infrequent, like the hook-like dark filament extending $>$7,300 km on the northern hemisphere in May 2 (Fig.~\ref{figure:IR1_clouds}B) and October 5 (Fig.~\ref{figure:IR1_clouds}J), suggesting the development of shear instabilities. A remarkable feature was also visible in October 10 (Fig.~\ref{figure:IR1_clouds}K), centered at 18$^{\circ}$S and with a size ranging 2,000-3,000 km. A candidate of wave packet with horizontal wavelength of 139 km was apparent during 2016 November 20 at $\sim$43$^{\circ}$N (see white frame and its zoom in Fig.~\ref{figure:IR1_clouds}I). Gravity waves were rare on NIR images during the VEx mission with one case found within 112 days of observations \citep[fig.~2C therein]{Peralta2008}, and Kelvin-type waves like the Y-feature \citep{Rossow1980,Kouyama2012,Peralta2015} are also missing. Since Kelvin-type waves require a stably stratified atmosphere, its absence in NIR imagery is consistent with the understanding that NIR images may be sensing cloud contrasts at an altitude range at which static stability is low \citep[fig.~5.9 therein]{Piccialli2010}.\\

\begin{figure*}[h!]
\centerline{\includegraphics[width=140mm]{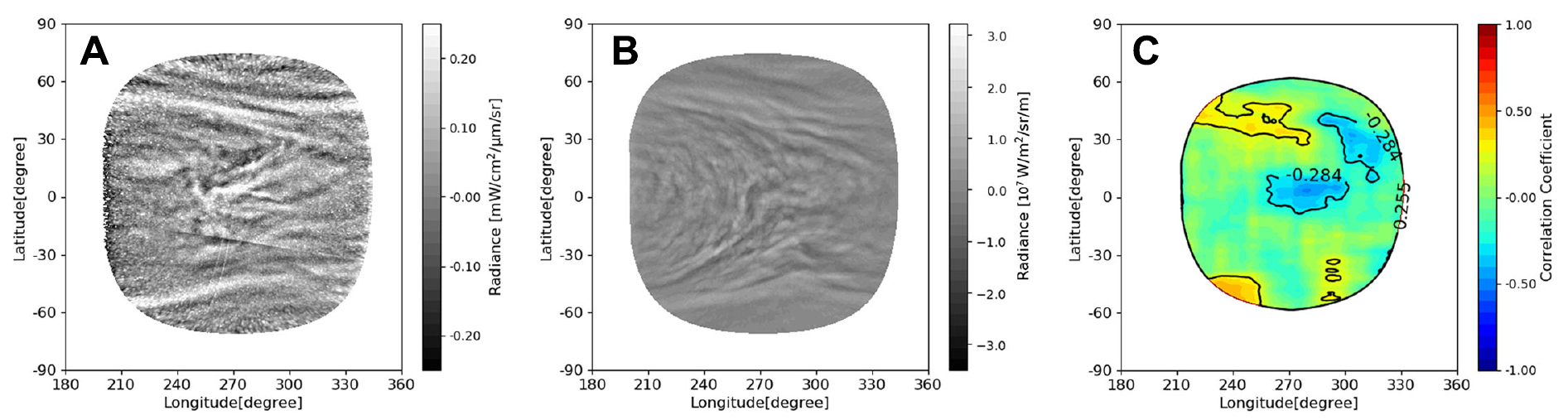}}
\caption{\textbf{Correlation between Venus's clouds at 900 and 365 nm.} The cylindrical projections of a pair of images acquired in 2016 May 17 at 08:02:07 by the camera IR1 (A) and at 08:17:18 by UVI (B) are compared to calculate a map of correlation between them (C). The projected images were processed with a Minnaert correction and a high-pass filtering by subtracting a Gaussian-smoothed image with FWHM of 6$^{\circ}$. The correlation map was obtained calculating the correlation coefficient pixel-to-pixel with a 24$^{\circ}\times$24$^{\circ}$ template. Meaningful values of correlation/anti-correlation were estimated to be +0.255/-0.284, these being estimated by comparing more than 30 pairs of uncorrelated IR1--UVI images, using an image from UVI image and another from IR1 acquired in different dates \citep{Narita2018}.}
\label{figure:IR1-UVI_Corr}
\end{figure*}

\null
During the Galileo flyby, \citet{Belton1991} reported anti-correlated cloud patterns on the albedo at lower latitudes of violet and NIR images, while during the VEx mission cases of positive correlation were found, suggesting that the unknown absorber might also affect NIR wavelengths \citep{Markiewicz2007b}. We studied the correlation between the cloud patterns on UVI 365-nm (cloud tops) and IR1 900-nm (middle clouds) radiance using cylindrical projections of Akatsuki Level-3 data \citep{Ogohara2017}. During the MESSENGER's flyby and the Akatsuki mission \citep{Peralta2017GRL,Limaye2018}, NIR and UV images seem uncorrelated in most of the cases. For instance, UV planetary-scale patterns like the Y-feature are absent in NIR images, while the sharp discontinuities in NIR (Fig.~\ref{figure:IR1_clouds}C) are missing in UV images. This supports the idea that contrast-forming processes may happen at different altitudes, as suggested by \citet{Belton1991}. In less frequent cases, the degree of correlation can be higher, as it can be observed in 2016 May 17 (see Fig.~\ref{figure:IR1-UVI_Corr}). The correlation between the spiral bands at high latitudes and the lack of correlation for the fine details at low latitudes is consistent with the weaker/stronger vertical wind shear at high/low latitudes \citep{Peralta2007b,Hueso2015}.

\section{Contrasts on the 900-nm albedo and implications}\label{sec:albedo}
Unlike the contrasts of up to 40\% reported for UV \citep{Belton1991,Lee2015Icar}, the NIR albedo during past missions exhibited weaker contrasts of $\sim$4\% \citep{Belton1991,Hueso2015,Khatuntsev2017}. We calculated the contrast of Venus's NIR albedo ($Contrast=100\cdot\left(I_{max}-I_{min}\right)/I_{max}$) using 519 photometrically corrected IR1 images acquired from 2015 December 7 to 2016 December 9 and solar zenith angles (SZA) and emission angles (EA) $<$60$^{\circ}$. Unexpectedly high contrasts were found, regardless of the variable distance (850--11,700 km) between maximum and minimum radiance. After applying the photometric correction (see subsection \ref{ssec:photometry}) and considering previous limitations for SZA and EA, contrasts between two extremes on the globe range 3\%--21\%, presenting an average and standard deviation of 13\%$\pm$3. Mean contrast decreases to 10\%$\pm$3 when considering only latitudes within 20$^{\circ}$N--20$^{\circ}$S, and local contrasts of up to 8\% can be observed in spatial scales $<$12$^{\circ}$.\\

\begin{figure}[h!]
\centerline{\includegraphics[width=80mm]{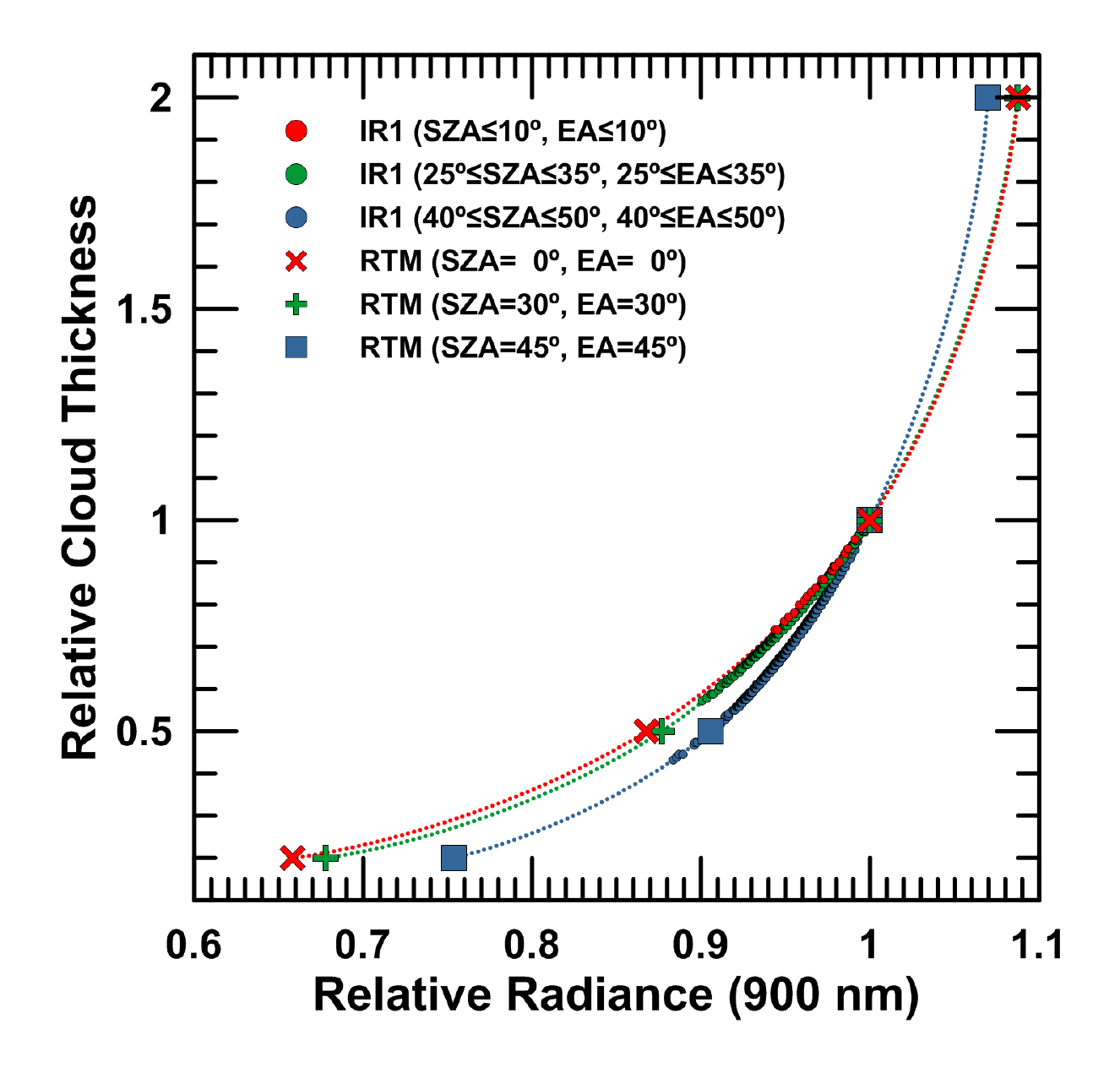}}
\caption{\textbf{Cloud thickness variations from contrasts in 900-nm photometrically-corrected radiance.} The relative cloud thickness as a function of the 900-nm radiance variations (crosses) was calculated for sets of solar zenith angle and emission angle [SZA,EA] using the radiative transfer model by \citet{Takagi2011}. Intermediate values between crosses were interpolated using spline fits (dotted lines). The interpolated curves for [SZA,EA]=0$^{\circ}$,30$^{\circ}$ and 45$^{\circ}$ (red,green,blue) were used to calculate the relative cloud thickness associated to the relative radiance in IR1 images (dots) measured for [SZA,EA] intervals 0$^{\circ}$--10$^{\circ}$,25$^{\circ}$--35$^{\circ}$ and 40$^{\circ}$--50$^{\circ}$, respectively.}
\label{figure:OptThick}
\end{figure}

If contrasts are not caused by absorbers, the 900-nm photometrically-corrected radiance mostly depends on the cloud thickness \citep{Takagi2011}, with thicker clouds being brighter and thinner clouds dimmer. This dependence is displayed for a combinations of SZA and EA in Figure \ref{figure:OptThick}, along with an approximate estimation of the change in cloud thickness associated to radiance contrasts from IR1 images for three combinations of ranges of [SZA,EA]. Most of the contrasts display values of relative radiance $>$0.9 ($<$10\%), implying changes of up to $\sim$40\% in the cloud thickness that were also considered by \citet{Takagi2011}. Elseway, spectra from VEx/SPICAV-IR \citep[fig.~15 therein]{Korablev2012} and MESSENGER/MASCS \citep[fig.~8 therein]{Perez-Hoyos2018} suggest the presence of some absorption bands that may be partially responsible for the higher contrasts observed at 900 nm. H$_{2}$O \citep{Cottini2012}, the controversial CH$_{4}$ \citep{Donahue1993} or even new candidates may be considered as potential absorbers to explain these contrasts.

\section{Winds from images at 900 nm and 1 $\mathrm{\mu m}$}\label{sec:winds}
A total of 511 wind measurements were obtained with cloud tracking using NIR images from Akatsuki/IR1 and, from the first time, using ground-based observations with NIR filters. Pairs of images were selected in order to maximize both the time and spatial coverage, providing winds for 43 days (to be extended in future works) from 2015 December 7 to 2017 January 13. On average, the error of individual measurements was about 7 m s$^{-1}$, ranging from 21 m s$^{-1}$ down to 0.6 m s$^{-1}$ in some specific cases when cloud tracers were stable enough to be unambiguously identified after 24 hours. The size of cloud tracers varies from 770 km to 4,100 km (in case of images from amateur observations), being selected depending on the spatial resolution, image signal-to-noise ratio and contrast of the observed patterns. We assume that zonal speeds do not dramatically depend on the size if the tracers, as shown for the nightside clouds \citep[fig.~7A therein]{Peralta2018ApJS}. The fast sharp discontinuities on the albedo (Fig.~\ref{figure:IR1_clouds}C) were interpreted as atmospheric waves and their speeds were discarded from this analysis.\\

\begin{figure*}
\centerline{\includegraphics[width=150mm]{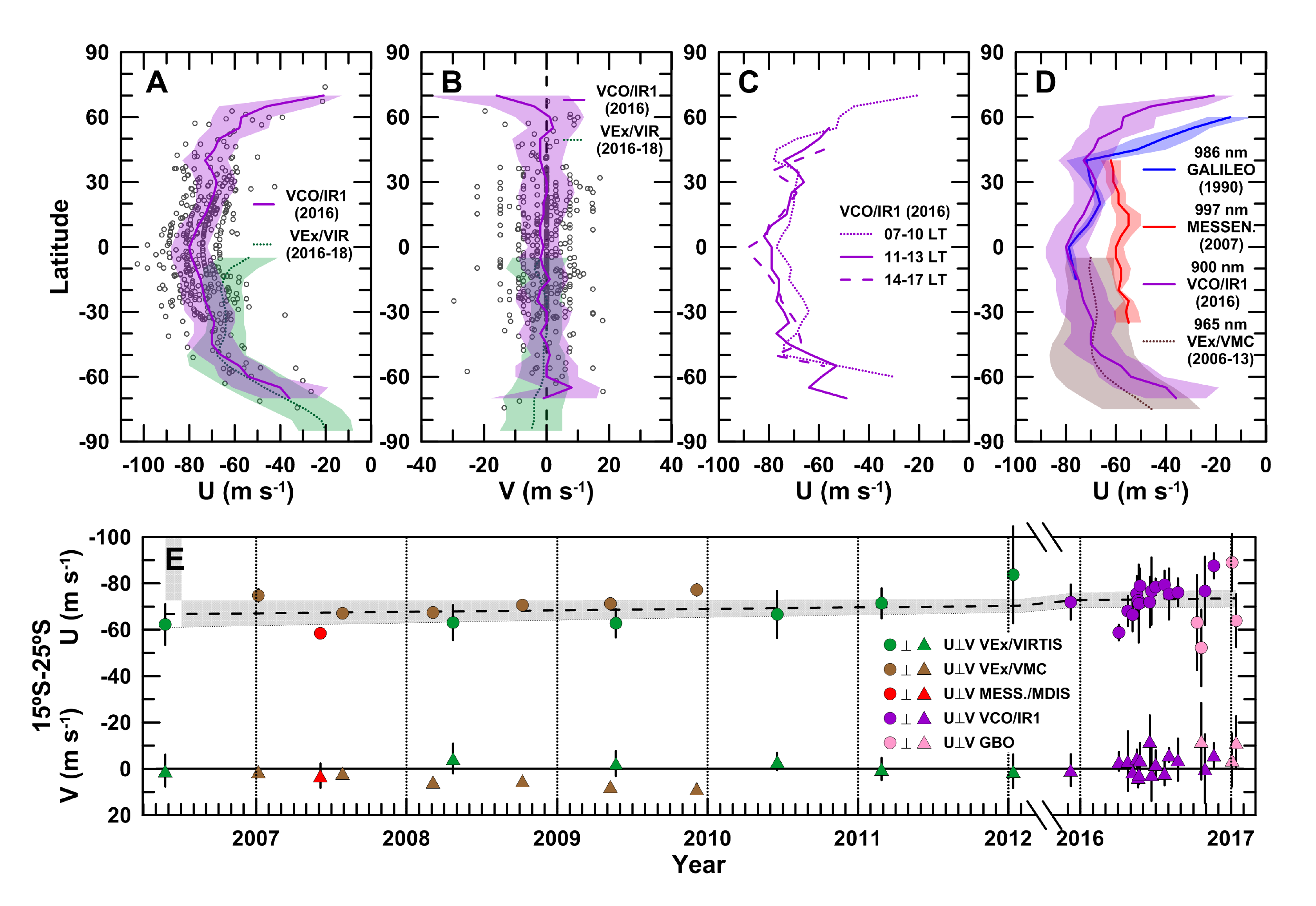}}
\caption{\textbf{Winds from Akatsuki/IR1 images and ground-based observations.} \textbf{(A)} and \textbf{(B)} show latitudinal profiles of zonal and meridional speeds zonally-averaged in latitude bins of 10$^{\circ}$ from Akatsuki (VCO) IR1 images (purple line) and VEx/VIRTIS (green line) \citep{Hueso2015}. Individual measurements from VCO/IR1 are also displayed (circles). \textbf{(C)} display zonal winds from IR1 zonally-averaged in intervals of local time. \textbf{(D)} compares profiles of zonal winds obtained in different years: flybys of NASA's Galileo (blue line) \citep{Peralta2007b} and MESSENGER (red) \citep{Peralta2017GRL}, VEx/VMC (brown) \citep{Khatuntsev2017} and VCO/IR1 (purple). Finally, \textbf{E} displays the long-term behaviour of zonal (dots) and meridional winds (triangles) within 15$^{\circ}$S--25$^{\circ}$S from VEx/VIRTIS (green colour), VEx/VMC (brown), MESSENGER/MDIS (red), VCO/IR1 (purple) and ground-based observations --GBO-- (pink). A linear fit and its confidence levels (95\%) is also displayed (dashed line).}
\label{figure:IR1_winds}
\end{figure*}

\null
The latitudinal profiles for the zonally-averaged winds at the middle clouds during 2016 is shown in Figs.~\ref{figure:IR1_winds}A and \ref{figure:IR1_winds}B. Although consistent at higher latitudes, zonal winds from IR1 peak at the equator and differ from the profile of constant zonal wind between the equator and midlatitudes reported in the past \citep{Belton1991,Sanchez-Lavega2008,Peralta2017GRL,Hueso2015}. A similar result was obtained for the zonal winds at the nightside lower clouds with IR2 images \citep[fig.~5A therein]{Peralta2018ApJS}, suspected to be originated by the sporadic jets forming at the equator \citep{Horinouchi2017NatGeo}. Confirming previous findings from VEx \citep{Hueso2015} no clear trend is observed on the meridional winds (Fig.~\ref{figure:IR1_winds}B) and the zonal winds seem to lack of a local time dependence (Fig.~\ref{figure:IR1_winds}C).\\

\null
A comparison among zonal winds' profiles in 1990 (Galileo's flyby), 2007 (MESSENGER's flyby), 2006--2013 (VEx/VMC and VEx/VIRTIS) and 2016--2017 (Akatsuki/IR1 and ground-based) (Fig.~\ref{figure:IR1_winds}D) suggests that the latitude where winds begin their poleward decay may be variable, and that speeds at lower latitudes can vary ($\sim$20 m s$^{-1}$) along the years. Differences in the vertical sensing for each filter might be another cause for these effects, although similar temporal changes were observed from VEx/VIRTIS images \citep[figs.~2a,3a therein]{Sanchez-Lavega2008}.\\

\null
Fig.~\ref{figure:IR1_winds}E shows zonal and meridional winds on the middle clouds during the MESSENGER flyby \citep{Peralta2017GRL} and the VEx and Akatsuki missions \citep{Hueso2015,Khatuntsev2017}. Data points from VEx/VIRTIS and Akatsuki/IR1 correspond to time averages of about 10 days or more, while those of VEx/VMC and MESSENGER/MDIS correspond to single days. To perform a coherent comparison with published data from VEx/VMC images at 965 nm, only winds between 15$^{\circ}$S and 25$^{\circ}$S were considered. Full data spans along 11 years and show that during the VEx mission zonal winds could vary in up to $\sim$15 m s$^{-1}$ for timescales of $\sim$6 months and even shorter ($\sim$3 months, \citealt{Sanchez-Lavega2008}) during the Akatsuki mission. Contrarily to zonal winds at cloud tops \citep{Hueso2012,Khatuntsev2013,Kouyama2013,Sanchez-Lavega2017}, no steady increase is apparent at middle clouds along the years. Mean meridional winds from VMC 965-nm images were reported to be polewards \citep{Khatuntsev2017}, although time averages on different time ranges from both VIRTIS \citep{Hueso2015} and IR1 display a variable sense of circulation with time.\\

\section{Conclusions}\label{sec:conclus}
The 900-nm images acquired by Akatsuki/IR1 during 2016 have revealed unexpected features of the middle clouds on the dayside of Venus, with cloud morphologies not previously observed --such as strong hemispherical asymmetries, sharp albedo discontinuities or long hook-like stripes-- subject to rapid changes and unrelated to the patterns at the cloud tops. The 900-nm albedo exhibits unexpected high contrasts ranging 3\%--21\%, what may be caused by a variation of up to 40\% in the optical thickness of the clouds when there are no absorbers. Finally, we provide measurements of the winds at the middle clouds using IR1 images and, for the first time, Earth-based observations. Mean zonal winds are found to weakly peak at the equator, while combined data from VEx and Akatsuki missions along 10 years reveals long-term variations of the zonal winds of up to 20 m s$^{-1}$.


\acknowledgments
J.P. acknowledges JAXA's International Top Young Fellowship (ITYF). N.I. acknowledges partial support by JSPS KAKENHI Grant Number JP16H02225. A.S.-L. and R.H. were supported by the Spanish MINECO project AYA2015-65041-P with FEDER, UE support and Grupos Gobierno Vasco IT-765-13. All authors acknowledge the members of the L3 team for the correction in the navigation of the Akatsuki images. Albedo calculations were performed on a Supermicro SuperServer Intel(R) Xeon(R) CPU E5-2620 v4 funded through JAXA's IYTF. We are also grateful to the anonymous reviewers for their useful comments to improve the manuscript.\\


\end{document}